\documentclass[twocolumn,twoside,slac_two]{revtex4}
\usepackage{graphicx}
\usepackage{fancyhdr}
\usepackage{amsmath}
\usepackage{amsfonts}
\begin{document}

\title{Kinks in two-dimensional Anti-de Sitter Space}


\author{J.~L.~Barnes, D.~P.~Schroeder, T.~ter Veldhuis, and M.~J.~Webster}
\affiliation{Department of Physics and Astronomy, Macalester College, Saint Paul, MN 55105, USA}

\begin{abstract}
Soliton solutions in scalar field theory defined on a two-dimensional Anti-de Sitter background space-time are investigated. It is shown that the lowest soliton excitation generically has frequency equal to the inverse radius of the space-time. Analytic and numerical soliton solutions are determined in \lq\lq phi to the fourth" scalar field theory with a negative mass-squared. The classical soliton mass is calculated as a function of the ratio of the square of the mass scale of the field theory over the curvature of the space-time. For the case that this ratio equals unity, the soliton excitation spectrum is determined algebraically and the one-loop radiative correction to the soliton mass is computed in the  semi-classical approximation.
\end{abstract}

\maketitle

\thispagestyle{fancy}


%%%%%%%%%%%%%%%%%%%%%%%%%%%%%%%%%%
\section{Introduction}
Solitons (kinks) are topologically stable localized solutions to the equation of motion with finite energy. They have been extensively studied in scalar field theories in two-dimensional Minkowski space\cite{Rajaraman:1982is,Rebbi:1985wg}. The main objective of the present work is to extend well-known results obtained  for solitons in flat Minkowski space to
two-dimensional Anti-de Sitter (AdS) space\cite{Hawking:1973uf}, which has constant negative curvature. 

After recalling some of the properties of AdS space and establishing conventions and notation in Sect.~\ref{sectwo}, some general features of soliton excitation spectra in AdS space that follow from symmetry considerations are discussed in Sect.~\ref{secthree}. In contrast to the situation in Minkowski space, the soliton excitation spectrum in AdS space is discrete. The frequency of the lowest excitation is generically equal to the inverse of the AdS radius. This lowest mode is obtained by acting with a broken $SO(2,1)$ symmetry generator on a static soliton solution. It is the AdS equivalent of the generic zero mode of solitons in Minkowski space related to the spontaneously broken translation symmetry. The definition of a soliton's mass in terms of the energy momentum tensor is provided as well.

In Sect.~\ref{secfour} a simple scalar field theory is presented that features two degenerate vacua. Explicit analytic soliton solutions interpolating between the vacua are determined  in Sect.~\ref{secfive} for some specific values of a dimensionless parameter $\alpha$, which is the ratio of the square of the mass scale of the field theory over the curvature. This parameter dictates the classical dynamics of the model. The mass of the analytic solitons is calculated as well. Some numerical soliton solutions for generic values of the parameter $\alpha$ are also presented.   

The scalar field theory is quantized by canonical quantization in Sect.~\ref{secsix}. It is then rendered finite by normal ordering of the Hamiltonian in the trivial sector. This procedure unambiguously fixes the finite parts of the counter terms. A mode number cut-off is employed as an ultra-violet regulator. This regulator explicitly breaks the $SO(2,1)$ symmetry and as a consequence a counter term is generated that is not invariant. 

In Sect.~\ref{secseven} the vacuum energies in the trivial and one-soliton sectors are  calculated in the case that the parameter $\alpha$ equals unity,  and the quantum corrections to the soliton mass are obtained in this case to one loop order in the semi-classical approximation. Some concluding remarks are reserved for Sect. \ref{seceight}.

\section{Two-dimensional Anti-de Sitter space \label{sectwo}}
$AdS_{1+1}$ space can be viewed as an $SO(2,1)$ invariant hyperboloidal hypersurface
\begin{eqnarray}
({X^0})^2 -({X^1})^2+({X^2})^2 &=& \frac{1}{m^2},
\end{eqnarray}
embedded in a three-dimensional pseudo-Euclidean space with invariant interval
\begin{eqnarray}
ds^2 &=& ({dX^0})^2 -({dX^1})^2+({dX^2})^2,
\end{eqnarray}
where $m$ parameterizes the inverse length scale in $AdS_{1+1}$ space.  Global coordinates $x$ and $t$ can be defined as
\begin{eqnarray}
X^0 &=& \frac{1}{m} \cos(mt)\cosh(mx) \nonumber \\
X^1 &=& \frac{1}{m} \sinh(mx) \nonumber \\
X^2 &=& \frac{1}{m} \sin(mt)\cosh(mx). 
\end{eqnarray}
The  induced metric on the hypersurface in these coordinates is
\begin{eqnarray}
\label{metric} ds^2 &=& g_{\mu\nu} dx^\mu dx^\nu = \cosh^2(mx)dt^2-dx^2.
\end{eqnarray}
The hyperboloidal surface is covered once by $mt \in \left[ -\pi,\pi \right]$ and $mx \in \left< -\infty,\infty\right>$. The space has topology $S$(time)$\times$ $R$(space). In order to avoid closed time-like curves the universal covering space is considered by unwrapping the $S$, and the restriction on the range of the time coordinate is lifted. Boundaries are located at $x \rightarrow \pm \infty$. Well defined time evolution requires specification of consistent boundary conditions in addition to initial conditions at a Cauchy surface\cite{Avis:1977yn}.

The geodesic trajectories in these coordinates can be obtained by acting with a finite $SO(2,1)$ transformation on a particular geodesic trajectory, for example $\sinh(mx)=0$, resulting in
\begin{eqnarray*}
&& \sinh(mx)  =  \nonumber \\
&& \frac{ \sinh(\eta_2) \sin(m\left[t-t_0\right])}{\sqrt{\cosh^2(\eta_2) \cos^2(m\left[t-t_0\right]) + \sin^2 (m\left[t-t_0\right])}},  \nonumber
\end{eqnarray*}
where $t_0$ and $\eta_2$ are parameters of a finite transformation that reflect the initial conditions.
For infinitesimal parameters $t_0$, $\eta_1$, and $\eta_2$ these transformations are realized on the coordinates $x$ and $t$ as
\begin{eqnarray}
mt' & = & mt - m t_0-\eta_1 \sin(mt) \tanh(mx) \nonumber \\
   &    & + \eta_2 \cos(mt) \tanh(mx), \nonumber\\
mx' & = & mx  +  \eta_1 \cos(mt) +\eta_2 \sin(mt).
\label{transformations}
\end{eqnarray}

\begin{figure}[ht]
\centering
\includegraphics[width=80 mm]{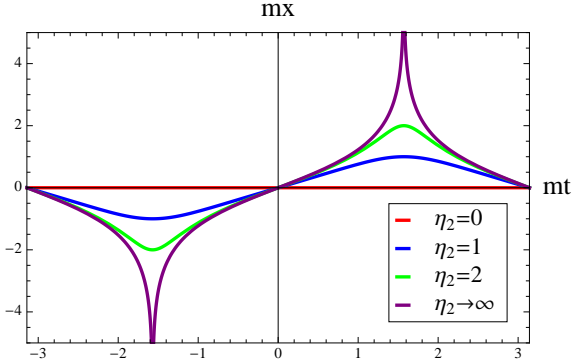} 
\caption{Projection of geodesic trajectories for $t_0=0$ and various values of $\eta_2$. The null geodesic corresponds to $\eta_2 \rightarrow \infty$ and is seen to reach the boundary in finite coordinate time. In this coordinate system the geodesics appear to oscillate around $x=0$ with frequency $\omega=m$ independent of amplitude.}
\label{geodesics}
\end{figure}
Null geodesics are obtained in the limit $\eta_2 \rightarrow \infty$. Figure (\ref{geodesics}) shows how various periodic geodesic trajectories, all for $t_0=0,$ start at one point, initially diverge and eventually converge again to a single point. Null geodesics are seen to reach the boundary of space-time in finite coordinate time. Massive particles never reach the boundary.

\section{Generalities \label{secthree}}

In order for stable kink solutions to exist in a scalar field theory it must feature two degenerate discrete vacua. This situation naturally occurs in models with a spontaneously broken $Z_2$ symmetry. For a theory with a single scalar field $\phi$ and canonical kinetic term the $SO(2,1)$ invariant Lagrangian density takes the form
 \begin{eqnarray}
\mathcal{L}  & =  & \frac{1}{2} g^{\mu\nu} \partial_\mu \phi \partial_\nu \phi -V(\phi).
\label{lagrangian}
\end{eqnarray}
A restricted class of potentials $V(\phi)$ is considered that have two degenerate minima at $\pm \phi_0$. A kink at rest is a static solution to the equation of motion and therefore satisfies the differential equation
\begin{eqnarray}
- \frac{d^2\phi}{d x^2} - m  \tanh(mx)\frac{d\phi}{d x}  
 + \frac{dV}{d\phi} & = & 0,
\label{solitoneq}
\end{eqnarray}
subject to the boundary conditions $\phi(x \rightarrow -\infty)= - \phi_0$ and $\phi(x \rightarrow \infty) =  \phi_0$. Note that the explicit appearance of the coordinate $x$ in this second order differential equation implies that it cannot be trivially integrated to yield a first order BPS equation. 

If  a kink solution $\phi_{\rm sol}(x)$ exists, then it is mapped onto another, time dependent solution to the equation of motion by spontaneously broken  $SO(2,1)$ transformations. For example, under a  transformation with infinitesimal parameter $\eta_1$ as defined in Eq.(\ref{transformations}) a static soliton solution transforms as
\begin{eqnarray}
\phi_{\rm sol}(x) & \rightarrow & \phi_{\rm sol} (x + \frac{\eta_1}{m} \cos(mt)) \nonumber \\
 & = & \phi_{\rm sol}(x) + \frac{\eta_1}{m} \frac{d \phi_{\rm sol}}{dx} \cos(mt).
\end{eqnarray}
Therefore, generically a kink in Anti-de Sittter space has a lowest excitation mode with frequency equal to $m$. This mode is a direct consequence of the spontaneously broken $SO(2,1)$ symmetry. It is the analogue of the zero mode of kinks associated with spontaneously broken translation symmetry in Minkowski space, and it corresponds to the kink following a periodic geodesic trajectory as discussed in Sect.~\ref{sectwo}.

The soliton mass is defined in terms of the conserved covariant energy momentum tensor, which is given by
\begin{eqnarray}
T_{\mu\nu} & = & \partial_\mu \phi \partial_\nu \phi - g_{\mu\nu} {\cal L}. 
\end{eqnarray}
The classical energy functional is 
\begin{eqnarray}
E[\phi ]& = & \int_{-\infty}^\infty dx \sqrt{-g}\, T^0_{\,\,\, 0}.
\label{functional}
\end{eqnarray}
As in Minkowski space, the classical soliton mass is defined to be  the difference in energy between the soliton and trivial vacuum field configurations as
\begin{eqnarray}
M_{sol} & = & E[\phi_{\rm sol}] - E[\phi_0] .
\label{solitonmasseq}
\end{eqnarray}
However, in the AdS$_{1+1}$ background Eq.(\ref{functional}) cannot be integrated to yield a BPS bound on the soliton mass.

\section{Explicit model \label{secfour}}

For simplicity, and to maintain a clear connection with well-studied solitons in Minkowski space,  a \lq\lq phi to the fourth" scalar field theory is considered with a negative mass-squared term, so that the potential reads
\begin{eqnarray}
V(\phi) &  = &  \frac{1}{2}(-\mu^2+\lambda\phi^2)^2. 
\end{eqnarray}
This potential is shown in the left panel of Fig.(\ref{solitonpic}).
In each of the two degenerate minima of the potential at $\phi_0 = \pm \mu/\sqrt{\lambda}$  the $Z_2$ symmetry $\phi \rightarrow -\phi$ is spontaneously broken. The maximum of the potential at $\phi=0$ also corresponds to a perturbatively stable vacuum as long as the dimensionless ratio $\alpha=2\lambda \mu^2/m^2$ is smaller than $1/4$ \cite{Dusedau:1985ue}. In terms of the dimensionless variables
\begin{eqnarray}
s & \equiv &  xm, \,\,\,\,\,\,\,\,\,\,\,\,\,\,\,\,\,  \tau \equiv tm \,\,\,\,\,\,\,\,\,\,\,\,\,\,\,\,\,  
\sigma  \equiv  \sqrt{\lambda}\frac{\phi}{\mu},  
\end{eqnarray}
the  Euler-Lagrange equation of motion  takes the form
\begin{eqnarray}
\frac{1}{\cosh^2(s)}\frac{\partial^2 \sigma}{\partial \tau^2} - \frac{\partial^2\sigma}{\partial s^2} - \tanh(s)\frac{\partial\sigma}{\partial s}  & & \nonumber \\
 + \alpha \sigma(-1+\sigma^2)  & = & 0. 
\label{eqofmotion}
\end{eqnarray}
The ratio $\alpha$, the square of the mass scale of the field theory over the curvature of the space-time background, is identified as the physical parameter that controls the classical dynamics.

%%%%%%%%%%%%%%%%%%%%%%%%%%%%%%%%%%
\section{Analytic soliton solutions \label{secfive}}
\begin{figure*}[ht]
\centering
$\begin{array}{cc} 
\includegraphics[width=75 mm]{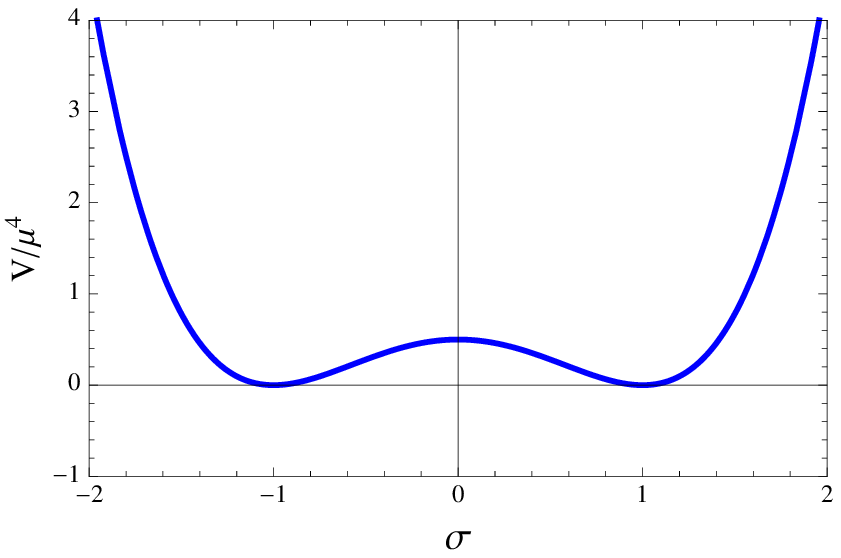} & \hspace{1 cm}
\includegraphics[width=75 mm]{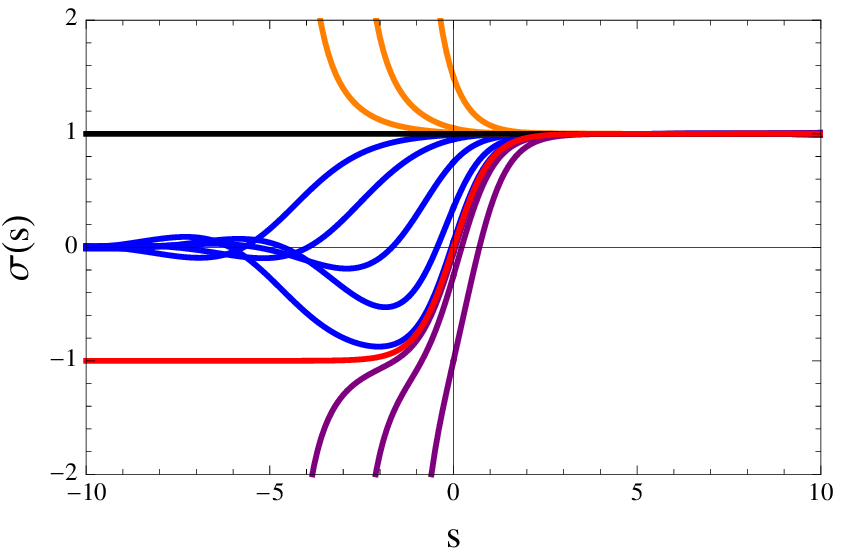}
\end{array}$
\caption{The left panel shows the scalar potential with two degenerate minima for the case $\alpha=1$. The right panel shows a family of static solutions to the equation of motion that satisfy the boundary condition $\sigma(s \rightarrow \infty)=1$, also for $\alpha=1$. Note that there exists only one solution which in addition satisfies the boundary condition $\sigma(s \rightarrow -\infty)=-1$.}
\label{solitonpic}
\end{figure*}

Analytic static soliton solutions are obtained for the following specific values of the parameter $\alpha$:
\begin{eqnarray}
\alpha=0:\,\,\,\,\,\,\,\, \sigma_{\rm sol}(s) & = & \frac{4}{\pi} \arctan\left[ \tanh \left(\frac{s}{2} \right) \right], \nonumber \\
\alpha=1:\,\,\,\,\,\,\,\, \sigma_{\rm sol}(s) & = & \tanh\left(s\right), \nonumber \\
\alpha \rightarrow \infty:\,\,\,\,\,\,\,\, \sigma_{\rm sol}(s) & = & \tanh \left( \sqrt{\frac{\alpha}{2}} s \right).
\label{analyticsolitons}
\end{eqnarray}
The soliton solution corresponding to $\alpha=1$ is depicted in the right panel of Fig.(\ref{solitonpic}) together with several other static solutions to Eq.(\ref{solitoneq}) for the same value of $\alpha$ that also satisfy the boundary condition $\sigma(s \rightarrow \infty)=1$. However, it is clear  from the diagram that there is only one unique solution which in addition satisfies the boundary condition $\sigma(s \rightarrow -\infty)=-1$. This contrasts with the situation in Minkowski space, where one static soliton solution can be translated over arbitrary distances to generate a one parameter family of static soliton solutions. 

\begin{figure*}[ht]
\centering
$\begin{array}{cc} 
\includegraphics[width=75 mm]{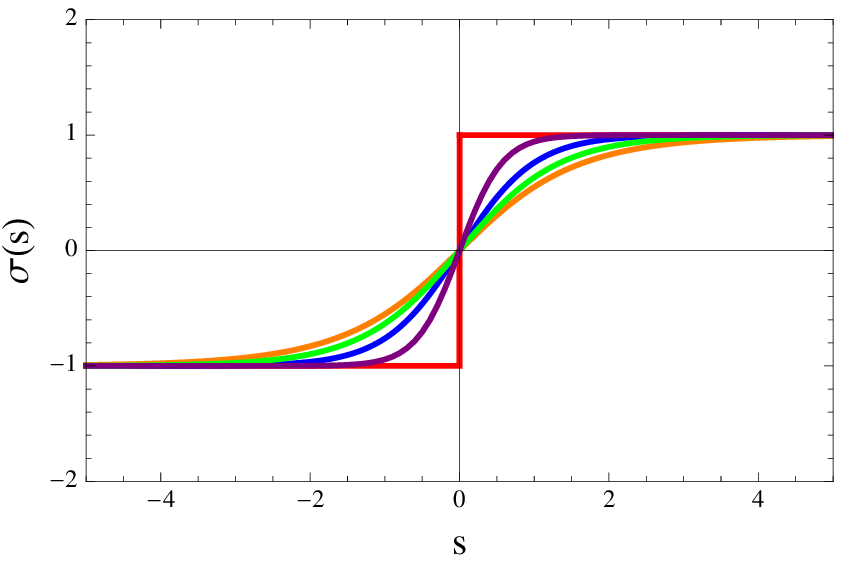} & \hspace{1 cm}
\includegraphics[width=75 mm]{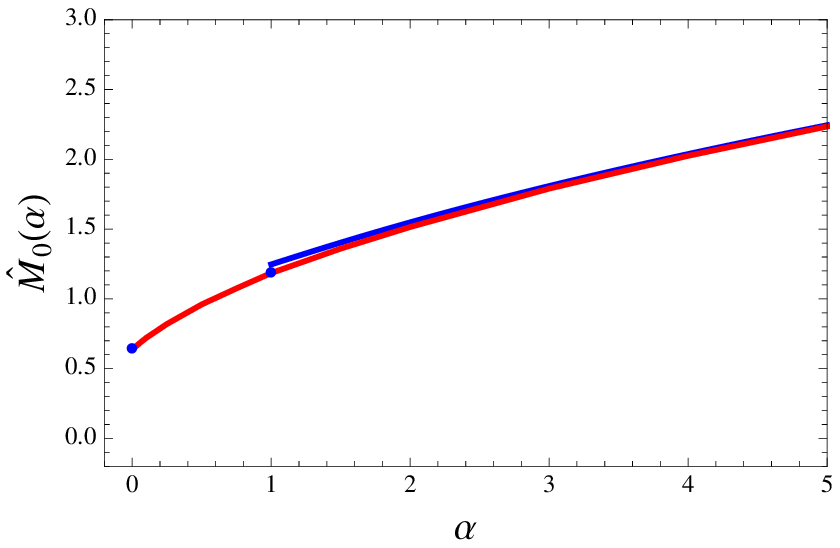}
\end{array}$ 
\caption{The left panel shows analytic soliton solutions for $\alpha=0$ (orange), $\alpha=1$ (blue), and $\alpha \rightarrow \infty$ (red), and numerical soliton solutions for $\alpha=0.25$ (green), and $\alpha=5$ (purple). The right panel displays the classical soliton mass as a function of the parameter $\alpha$. Analytic results for $\alpha=0$, $\alpha=1$, and $\alpha \rightarrow \infty$ are indicated in blue, while numerical results correspond to the red curve. }
\label{soliton-energy}
\end{figure*}

Even though they are not obtained in analytic form, soliton solutions also exist for generic values of $\alpha$. The left panel of Fig.(\ref{soliton-energy}) shows the analytic soliton solutions presented in Eq.(\ref{analyticsolitons}) together with numerical soliton solutions for two other representative values of $\alpha$, namely $\alpha=0.25$ and $\alpha=2.0$.

The masses of the solitons corresponding to the analytic solutions presented in Eq.(\ref{analyticsolitons}) are calculated according to Eq.(\ref{solitonmasseq}) to be
\begin{eqnarray}
\alpha=0:\,\,\,\,\,\,  \hat{M}_{0}  & = & \frac{2}{\pi}, \nonumber\\
\alpha=1:\,\,\,\,\,\, \hat{M}_{0}  & = & \frac{3}{8} \pi, \nonumber\\
\alpha \rightarrow \infty:\,\,\,\,\,\,\hat{M}_{0} & = & \frac{2}{3} \sqrt{2 \alpha} +\frac{(\pi^2 -6)}{18} \sqrt{\frac{2}{\alpha}},
\end{eqnarray}
where $M_{\rm sol}=( \mu^2 m/  \lambda) \hat{M}_{0}$. The masses of solitons for generic values of $\alpha$ are calculated numerically. Both the analytic and numerical results are displayed in the right panel of Fig.(\ref{soliton-energy}).

A time dependent solution representing a moving soliton is obtained by acting with a finite  $SO(2,1)$ transformation on a static solution. For example, defining the function
\begin{eqnarray}
sh(x,t) & = &\cosh(\eta_2) \sinh( mx)  \nonumber \\
     &    &+ \sinh(\eta_2) \cosh(mx) \sin(m\left[t-t_0\right]),\,\,\,\,\,\,\,\,\,
\end{eqnarray}
a time dependent solution to the equation of motion for $\alpha=1$ representing a soliton following a geodesic trajectory  takes the form
\begin{eqnarray}
\phi_{\rm sol} (x,t;\eta_2,t_0)& = & \frac{sh(x,t)}{\sqrt{1 + sh^2(x,t)}}.
\end{eqnarray}

\section{Quantization, regularization, and renormalization \label{secsix}}
\begin{figure}[ht]
\centering
\includegraphics[width=80 mm]{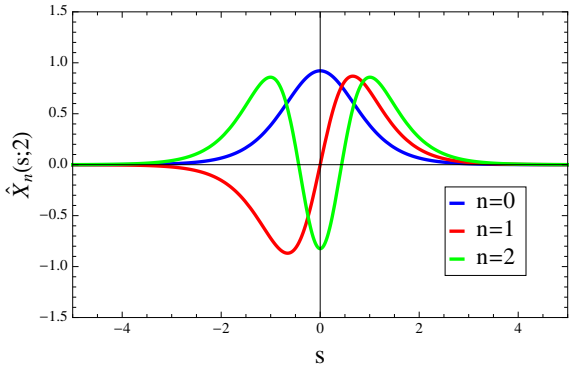} 
\caption{Normal mode functions for $g=2$.}
\label{modes}
\end{figure}

The scalar field theory is renormalized in the trivial (no-soliton) sector. This procedure also renders physical observables in the one-soliton sector finite.  In order to find the excitation spectrum in the trivial sector,  the scalar field is expanded around one of the equivalent degenerate vacua as
\begin{eqnarray}
\sigma(s,\tau) & = & 1 + \eta(s,\tau).
\end{eqnarray}
For small fluctuations only linear terms in $\eta$ need to be considered. In this approximation the equation of motion becomes
\begin{eqnarray}
 \frac{1}{\cosh^2(s)}\frac{\partial^2\eta}{\partial \tau^2}-\frac{\partial^2\eta}{\partial s^2}-\tanh(s)\frac{\partial\eta}{\partial s}+2 \alpha \eta &  = & 0.
\end{eqnarray}
This equation is solved by separation of variables. After making the Ansatz
\begin{eqnarray}
\eta(s,\tau) & = & e^{i \hat{\omega} \tau} \hat{X}(s),
\end{eqnarray}
the space dependent factor $\hat{X}(s)$ of  a normal mode is a solution to the equation
\begin{eqnarray}
-\cosh^2(s) \frac{d^2\hat{X}}{ds^2} - \sinh(s) \cosh(s) \frac{d\hat{X}}{ds}  & & \nonumber \\ 
+ 2 \alpha \cosh^2(s)\hat{X} & = &  \hat{\omega}^2\hat{X} . \,\,\,\,\,
\label{normmodes}
\end{eqnarray}
This equation has the form of a time-independent Schr{\"o}dinger equation, and familiar tools from quantum mechanics can be used  to  find the normalizable eigenfunctions and eigenvalues. The equivalent quantum mechanical model is of the Scarf I type, and therefore supersymmetry \cite{Witten:1981nf} and shape invariance \cite{Cooper:1994eh}can be employed to find the energy spectrum of the equivalent quantum mechanical model in a purely algebraic manner. The resulting soliton frequency spectrum takes the form
\begin{eqnarray}
\hat{\omega}_n & = & n+g, \,\,\,\,\,\,\,\,\,  n \in \mathbb{N}, 
\end{eqnarray}
with 
\begin{eqnarray}
g & = & \frac{1}{2} + \sqrt{\frac{1}{2}+2 \alpha}.
\end{eqnarray}
This evenly spaced spectrum is consistent with the unbroken $SO(2,1)$ symmetry in this vacuum \cite{Dusedau:1985ue}.
The normal mode functions can be written in terms of the Jacobi polynomials as
\begin{eqnarray}
\hat{X}_n(s;g) & = & B_n \frac{1}{\cosh^{g}(s)} P_n^{(g-\frac{1}{2},g-\frac{1}{2})}\left[\tanh(s)\right], \,\,\,\, 
\end{eqnarray}
where the normalization constant $B_n$ is chosen as
\begin{eqnarray}
B_n & = & \sqrt{ \frac{(n+g)}{2^{2 g-1}} \frac{\Gamma(n+1) \Gamma(n+2 g)}{\Gamma(n+g+\frac{1}{2})^2}},
\end{eqnarray}
so that the normal mode functions satisfy the orthonormality condition
\begin{eqnarray}
\int_{-\infty}^{\infty}  \frac{1}{\cosh s} \hat{X}^{\dagger}_n (s;g) \hat{X}_m(s;g) ds & = & \delta_{nm}. 
\end{eqnarray}
The first three normal mode functions are displayed in Fig.(\ref{modes}) for $g=2$.

After shifting the field, $\phi=\mu/\sqrt{\lambda}+\phi'$, the Hamiltonian density of the model is given by
\begin{eqnarray}
{\cal H} & = & \frac{1}{2} \pi^{\prime 2} + \frac{1}{2} \frac{\partial \phi'}{\partial x}^2 +V(\phi'),
\end{eqnarray}
where the canonical momentum $\pi'$ is defined as
\begin{eqnarray}
\pi' & = & \sqrt{-g} \frac{\partial \cal{L}}{\partial (\partial_0 \phi')} = \frac{1}{\cosh(mx)} \frac{\partial \phi'}{ \partial t}.
\end{eqnarray}
Both the field $\phi'$ and its canonical momentum $\pi'$ are expanded in terms of normal modes according to
\begin{eqnarray}
\phi' & = & \sum_{n=0}^N \frac{a_n}{\sqrt{2 \omega_n}} X_n(x;g) e^{i \omega_n t}  
\nonumber \\
& & \,\,\,\,\,\,+ 
\frac{a^\dagger_n}{\sqrt{2 \omega_n}} X^\dagger_n(x;g) e^{-i \omega_n t}, \nonumber 
\end{eqnarray}
and
\begin{eqnarray}
\pi' & = & \sum_{n=0}^N - i a_n \sqrt{\frac{\omega_n}{2}} \frac{X_n(x;g)}{\cosh(mx)} e^{i \omega_n t}  
\nonumber \\
 & & \,\,\,\,\,\,\,\,\,\,+ 
i a^\dagger_n \sqrt{\frac{\omega_n}{2}} \frac{X^\dagger_n(x;g)}{\cosh(mx)} e^{-i \omega_n t},
\end{eqnarray}
where $X_n(x;g) = \sqrt{m} \hat{X}_n(mx;g)$. In order to quantize the model, the expansion coefficients $a_n$ and $a_n^\dagger$ are promoted to operators \cite{Avis:1977yn} and the commutation relation
\begin{eqnarray}
\left[ a_n , a^\dagger_m \right] & = & \delta_{nm}
\end{eqnarray}
is imposed.

The theory is rendered finite by normal ordering. The normalization conditions are implied by this procedure and the finite parts of counter terms are therefore unambiguously fixed. The relation between the normal ordered Hamiltonian and the original Hamiltonian is
\begin{eqnarray}
: H: & = &  H - \delta D - \int_{-\infty}^{\infty} \cosh(mx) dx \left[ \delta \mu^2\lambda \phi^2  \right]. \nonumber
\end{eqnarray}
The mode sums are cut off at large mode number $N$ in order to regulate the theory in the ultra violet. The mass counter term is found to be logarithmically divergent,
\begin{eqnarray}
\delta \mu^2 & = & 3 \lambda  \sum_{i=0}^{N} \frac{1}{2 \omega_i} X_i(x;g) X_i^\dagger(x;g)  \\
 & = & \frac{3}{2 \pi} \lambda  \left( \ln(N) +   \ln(2)-\psi(g)  -\ln \left[ \cosh(mx)  \right] \right). \nonumber
\end{eqnarray}
The finite part of the mass counter term is seen not to be $SO(2,1)$ invariant through its explicit coordinate dependence. This is a consequence of the use of a regularization scheme that breaks the $SO(2,1)$ invariance explicitly.
The vacuum energy counter term has a quadratic divergence and reads
\begin{eqnarray} 
\delta D & = & \sum_{i=0}^{N} \frac{1}{2} \omega_i = \frac{1}{4} m (N+2 g) (N+1).  
\end{eqnarray}

\section{Quantum corrections to the soliton mass \label{secseven}}

By dimensional analysis, the soliton mass can be expanded as
\begin{eqnarray}
M_{\rm sol}(\alpha) & = & \frac{\mu^2 m}{\lambda} \Bigl[ \hat{M}_0(\alpha) + \hat{M}_1(\alpha) \left(\frac{\lambda \hbar}{\mu^2}\right) \nonumber \\
& & + \hat{M}_2(\alpha) \left(\frac{\lambda \hbar}{\mu^2}\right)^2 + \cdots \Bigr],
\end{eqnarray}
where the classical contribution given by $\hat{M}_0(\alpha)$ was discussed in Sect.~\ref{secfive} and the functions $\hat{M}_n(\alpha)$ for $n \geq 1$ reflect quantum corrections. The one-loop quantum correction can be determined analytically for $\alpha=1$ in the semi-classical approximation.
In order to find the excitation spectrum in the one-soliton sector for this value of $\alpha$, the scalar field is expanded around the classical soliton solution as
\begin{eqnarray}
\sigma(s,\tau) & =  & \tanh(s) + \eta(s,\tau). \nonumber
\end{eqnarray}
The linearized equation of motion for the field $\eta$ takes the form
\begin{eqnarray}
\frac{1}{\cosh^2(s)}\frac{\partial^2\eta}{\partial \tau^2}-\frac{\partial^2\eta}{\partial s^2}-\tanh(s)\frac{\partial\eta} {\partial s}& &
\nonumber \\
+3\tanh^2(s)\eta-\eta & = & 0. 
\end{eqnarray}
This equation is  solved by separation of variables according to the Ansatz
\begin{eqnarray}
\eta(s,\tau) & = & e^{i \hat{\omega} \tau} X(s). \nonumber
\end{eqnarray}
The space dependent part of the normal mode equation reads
\begin{eqnarray}
-\cosh^2(s) \frac{d^2X}{ds^2} - \sinh(s) \cosh(s) \frac{dX}{ds}  & & \nonumber \\
+ 2  \cosh^2(s)X -3X& = &  \hat{\omega}^2X ,\,\,\,\,\,\,
\end{eqnarray}
and is again equivalent to a Schr{\"{o}}dinger equation with a potential of the Scarf I type. The  soliton  excitation spectrum is determined to take the form
\begin{eqnarray}
\hat{\omega}_n^{\rm sol} & = & \sqrt{(n+2)^2 - 3}.  \,\,\,\,\,\,\,\,\,  n \in \mathbb{N}.
\end{eqnarray}
The lowest excitation frequency in the one-soliton sector, the one corresponding to $n=0$, is seen to be $\hat{\omega}_0^{\rm sol}=1$, as expected on general grounds from the spontaneous $SO(2,1)$ symmetry breaking discussed in Sect.~\ref{secthree}.

To one-loop order there are two contributions to the quantum corrections to the soliton mass \cite{Dashen:1974cj}. The first contribution is due to the mass renormalization and takes the form
\begin{eqnarray}
\Delta M_A & = & -   \lambda \int_{-\infty}^{\infty} cosh(mx) dx\, \delta\mu^2(x) \left[ \phi_{\rm sol}^2(x) - \phi_0^2 \right] \nonumber \\
 & = & \frac{3}{4} m \left[ \ln (N) +\gamma -1  \right],
 \end{eqnarray}
while the second contribution is the difference in vacuum energy between the trivial and one soliton sectors,
 \begin{eqnarray}
 \Delta M_B & = & \sum_{i=0}^{N}\frac{1}{2} \omega_i^{\rm sol} - \delta D \nonumber \\
  & = & -\frac{3}{4} m \left[ \ln (N) + \gamma -1 \right] + \frac{1}{2} m  C_S.
  \end{eqnarray}
Here Schroeder's number $C_S$ is finite and defined through the sum
\begin{eqnarray}
 C_S & = & \sum_{i=0}^{\infty}  \left[ \sqrt{(i+2)^2 -3} - (i+2) + \frac{3}{2} \frac{1}{(i+2)} \right] \nonumber \\
 &   \approx & -0.3485. 
\end{eqnarray}
The logarithmically divergent parts of the two contributions $\Delta M_A$ and $\Delta M_B$ cancel against each other. The finite, physical mass of the soliton including its one-loop quantum corrections is thus
\begin{eqnarray}
M_{\rm sol} & = & M_{\rm clas} +M_A+M_B=\frac{3 \pi}{16} \frac{m^3}{\lambda^2}   +  \frac{1}{2} m  C_S \nonumber \\
& = & \frac{\mu^2 m}{\lambda} \Bigl[ \frac{3 \pi}{8} + \frac{1}{2} C_S \left( \frac{\lambda \hbar}{\mu^2} \right) + \cdots \Bigr].
\end{eqnarray}
This result is valid for small values of $\lambda$ and $\alpha=1$.

\section{Discussion \label{seceight}}
A zero mode is generically obtained in the excitation spectrum of solitons in Minkowski space due to the spontaneously broken translation symmetry. In Anti-de-Sitter space, the corresponding spontaneously broken symmetry generator gives generically rise to a mode with $\omega=m$.
The supersymmetric and shape invariant quantum mechanical models  encountered in relation to the excitation spectra in the model are equivalent by a coordinate transformation to the Scarf I potentials \cite{Cooper:1994eh}.
The soliton solution in Anti-de Sitter space was obtained as a static solution to the second order Euler-Lagrange equation of motion. Due to the explicit coordinate dependence of the Hamiltonian density it is not possible to derive a BPS bound and a first order BPS equation as is done in Minkowski space. The interpretation of the mass of the soliton in Anti-de Sitter space as a topological charge seems unclear.

\begin{acknowledgments}
This work  was supported in part by a Cottrell Award from the Research Corporation and by the NSF under grant PHY-0758073. TtV thanks Donald Spector and Thomas Clark for interesting discussions.
\end{acknowledgments}

\bigskip % extra skip inserted

\end{document}